\documentclass[aps,prl,twocolumn,groupedaddress]{revtex4}

\usepackage{amsmath}
\usepackage{amssymb}
\usepackage{graphics}
\usepackage[dvips]{graphicx}
\graphicspath{{./},{./pictures/}}
\begin{document}

\title{Turbulence without inertia in quantum fluids}

\author{Demosthenes Kivotides}
\affiliation{Low Temperature Laboratory, Helsinki University of Technology,\\
P.O. Box 2200, FIN-02015 HUT, Finland}

\date{\today}

\begin{abstract}
Numerical calculations of $^{4}He-II$ hydrodynamics
show that a dense tangle of superfluid vortices
induces in an initially stationary normal fluid a highly dissipative,
complex, vortical flow pattern ("turbulence") with $k^{-2.2}$ energy 
spectrum scaling and fluctuations Reynolds number of 
order unity. In this normal fluid flow the effects of 
mutual friction excitation from the superfluid vortices and those of
viscous stresses are of the same order. The results suggest that 
in previous experiments the dynamics of decaying, high Reynolds 
number, quantum turbulent flows could only weakly be affected by the
quantized vortices. As a consequence, their energy spectra
would be (to a very good approximation) the classical, Navier-Stokes type,
energy spectra of the normal fluid component.
\end{abstract}

\pacs{67.40.Vs, 47.27.Ak, 47.27.Gs}

\maketitle

In quantum fluid turbulence \cite{vinen:2002}, a tangle of quantized vortices 
interacts via mutual friction forces 
with thermal excitations (known as normal fluid) of the superfluid ground state.
Until recently, theoretical investigations
involved only the dynamics of the superfluid vortices \cite{schwarz:1985}. 
The normal fluid flow was assumed to possess infinite inertia
and it was prescribed in a kinematic way.
Numerical and computational methods for 
fully dynamical quantized turbulence calculations were developed
for the first time in \cite{idowu:2001}.
Subsequently, the strategy of gradually increasing complexity was adopted
in performing a series of three dimensional calculations. 
First, it was shown in \cite{kivotidess:2000} that a superfluid 
vortex ring induces in an initially stationary 
normal fluid two coaxial vortices which together with the quantized ring
propagate as a triple ring structure.
A second calculation \cite{kivotidesr:2001} allowed a small number of reconnections
in order to show that Kelvin waves excited by the latter induced
a dramatic increase of kinetic energy dissipation rate in the normal fluid.
The present Letter introduces the element of vortex line density characteristic
of experiments while keeping the condition of initial stationarity for the
normal fluid.\\

The numerical method is described in detail in the mentioned references.
If $\boldsymbol{S}(\xi,t)$ is the three dimensional representation of the vortex tangle
(where $\xi$ is the arclength parametrization along the loops), 
then its motion obeys the equation \cite{idowu:2000}:
\begin{eqnarray}
\frac{d\boldsymbol{S}}{dt} = \boldsymbol{V_l}=
&&h \boldsymbol{V_s}+ h_{\times} \boldsymbol{S}^{\prime} 
\times (\boldsymbol{V_n}-
\boldsymbol{V_s})- \nonumber\\
&&h_{\times \times} \boldsymbol{S}^{\prime} \times (\boldsymbol{S}^{\prime}
\times \boldsymbol{V_n})
\end{eqnarray}
where the superfluid velocity $\boldsymbol{V_s}$ is given by the Biot-Savart integral:
\begin{equation}
\boldsymbol{V_s}(\boldsymbol{x}) \equiv 
\boldsymbol{V_i}(\boldsymbol{x})= \frac{\kappa}{4 \pi} \int
\frac{(\boldsymbol{S}-\boldsymbol{x}) \times
d\boldsymbol{S}}
{{|\boldsymbol{S}-\boldsymbol{x}|}^3}
\end{equation}
$t$ is time, $\boldsymbol{x}$ is space, $\kappa$ is the quantum of
circulation, $\boldsymbol{V_n}$ is the velocity
of the normal fluid, $\boldsymbol{S}^{\prime}=
\frac{d\boldsymbol{S}}{\|d\boldsymbol{s}\|}$ is the unit tangent vector while $h$,
$h_{\times}$ and $h_{\times \times}$ are constants related to mutual friction physics.
In writing equation $(1)$ one ignores the inertia of the vortices 
and the spectral-flow (Kopnin) force that
is relevant only for fermionic quantum fluids. The equation 
includes the Magnus, drag 
and Iordanskii forces \cite{idowu:2000}.
We call the sum of the Iordanskii and drag forces 
mutual friction.
In the definition of $\boldsymbol{V_s}$ any irrotational ground state motion
has been neglected.\\
\begin{figure*}[t]
\begin{minipage}[t]{0.79\linewidth}
\begin{tabular}[b]{c c}
\includegraphics[width=0.49\linewidth]{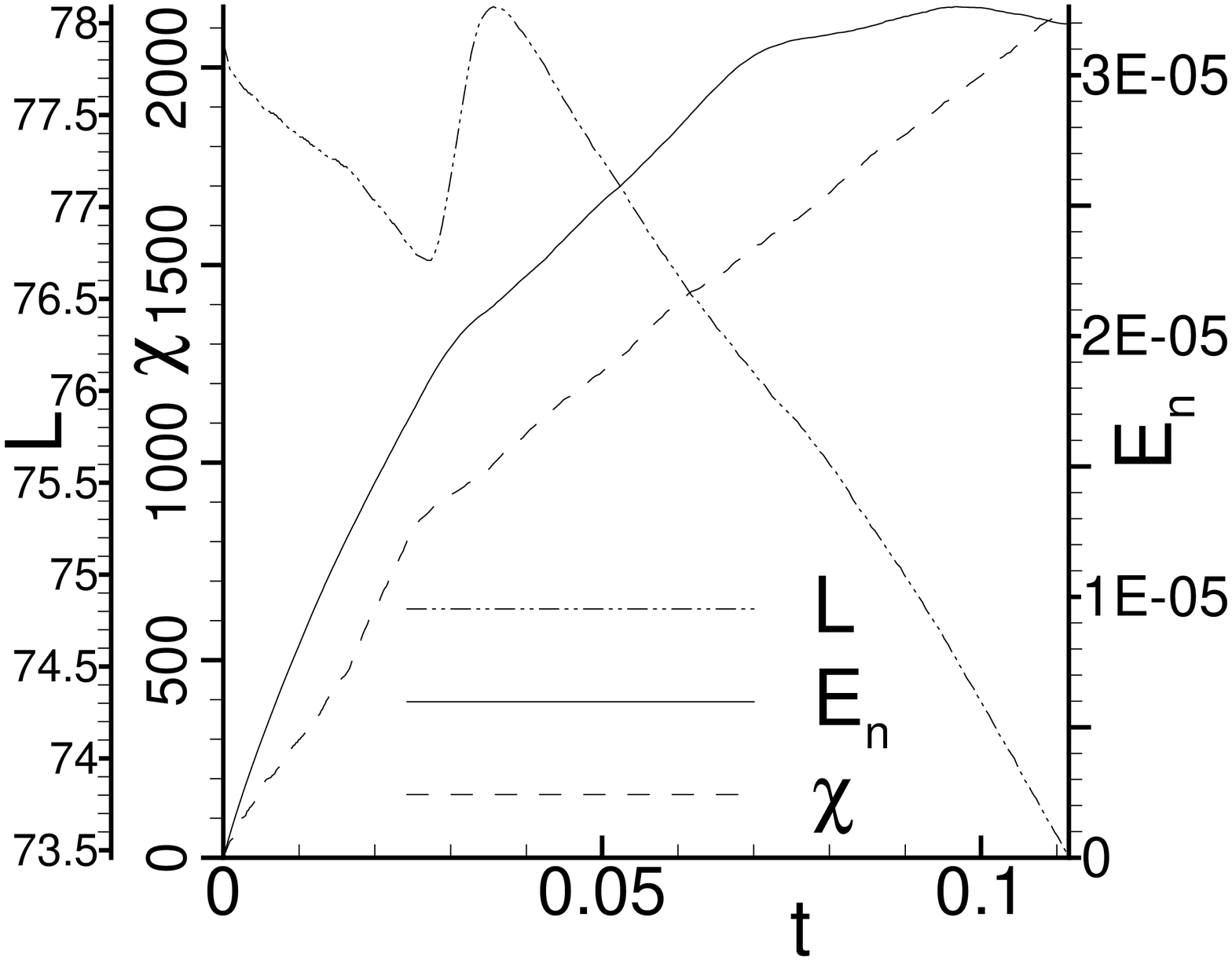}&
\includegraphics[width=0.49\linewidth]{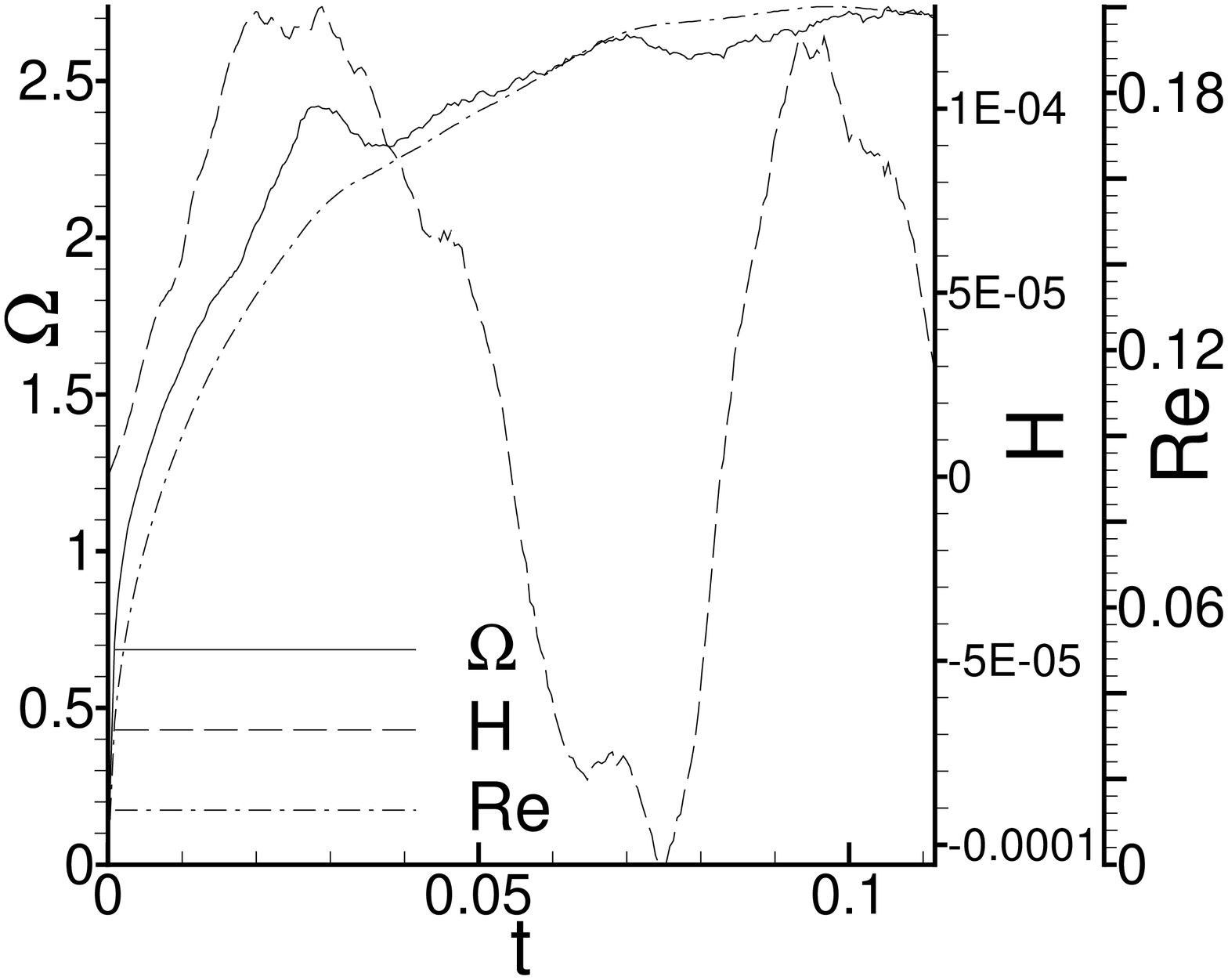}\\
\end{tabular}
\caption{\label{diagsone} Evolutions of: tangle length $L$, normal fluid
energy $E_n$, reconnections number $\chi$ (left); normal fluid enstrophy $\Omega$,
normal fluid helicity $H$ and normal fluid Reynolds number $Re$ (right).}
\end{minipage}
\end{figure*}
\begin{figure*}[t]
\begin{minipage}[t]{0.79\linewidth}
\begin{tabular}[b]{c c}
\includegraphics[width=0.49\linewidth]{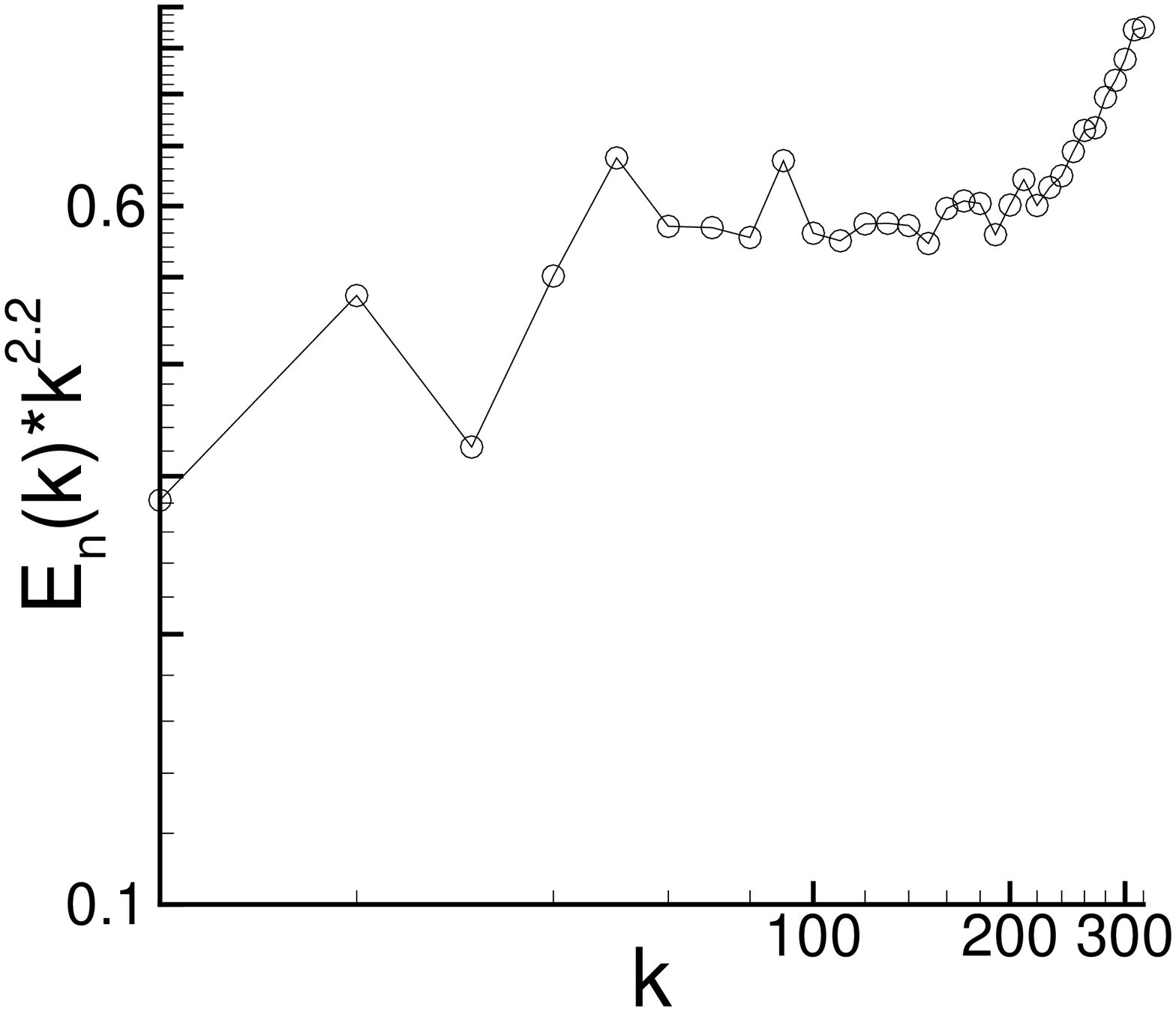}&
\includegraphics[width=0.49\linewidth]{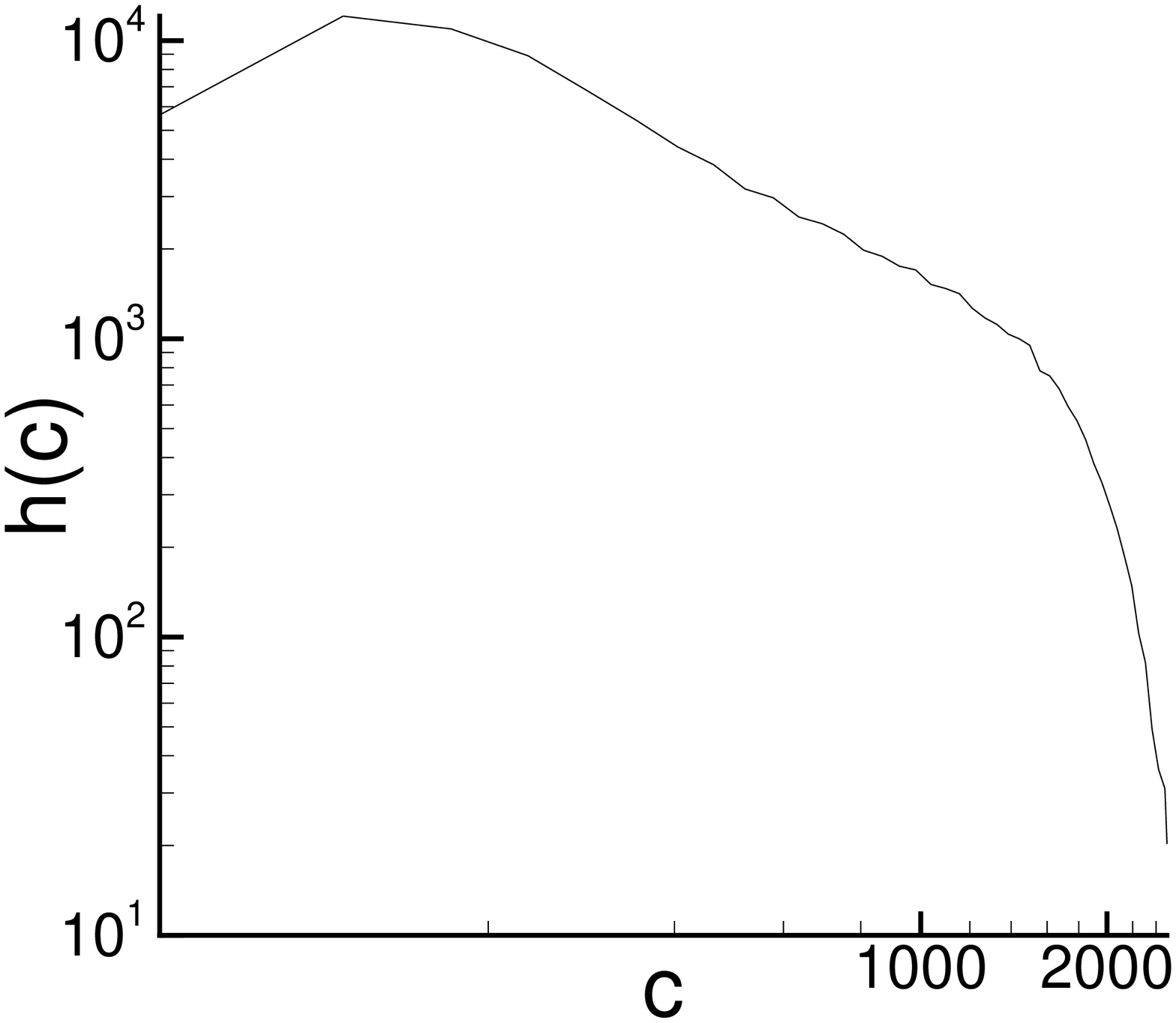}\\
\end{tabular}
\caption{\label{diagstwo} Normal fluid energy spectrum $E_{n}(k)$ at normal
fluid kinetic energy peak multiplied
by $k^{2.2}$ (left); histogram $h(c)$ of curvature of superfluid vortices at normal
fluid kinetic energy peak (right).}
\end{minipage}
\end{figure*}
The motion of the normal fluid is governed by the forced 
incompressible Navier-Stokes equation:
\begin{eqnarray}
\frac{\partial \boldsymbol{V_n}}{\partial t}+(\boldsymbol{V_n} 
\cdot \boldsymbol{\nabla}) \boldsymbol{V_n}=
-\frac{1}{\rho} \boldsymbol{\nabla} p +
\nu \nabla^2\boldsymbol{V_n}+\frac{1}{\rho_n}
\boldsymbol{F}
\end{eqnarray}
\begin{equation}
\boldsymbol{\nabla} \cdot \boldsymbol{V_n}=0
\end{equation}
where $\rho_n$ is the density of the normal fluid, 
$\rho$ is the
total density of the fluid, $p$ is the pressure, $\nu = \frac{\mu}{\rho_n}$
is the kinematic viscosity ($\mu$ stands for the dynamic viscosity) and $\boldsymbol{F}$
is the mutual friction force per unit volume.
The latter is being calculated from the formula 
for the sum of the drag and Iordanskii 
forces per unit vortex length $\boldsymbol{f}$:
\begin{eqnarray}
\boldsymbol{f}= &&\rho_s \kappa d_{\times \times} 
\boldsymbol{S}^{\prime} \times (\boldsymbol{S}^{\prime}
\times (\boldsymbol{V_n}- \boldsymbol{V_l}))- \nonumber\\
&&\rho_s \kappa d_{\times} \boldsymbol{S}^{\prime} \times (\boldsymbol{V_n}-
\boldsymbol{V_l}).
\end{eqnarray}
where $\rho_s$ is the density of the super fluid and 
the dimensionless term $d_{\times}$ incorporates both
the Iordanskii and the corresponding drag force component coefficients.
The numerical procedure for obtaining 
$\boldsymbol{F}$ from $\boldsymbol{f}$ 
detects first all segments of the vortex tangle inside a numerical grid cell.
Then, it numerically integrates the known $\boldsymbol{f}$ over 
the length of these segments
and divides with the cell volume. Grid cells that contain no vortices
have zero mutual friction force.\\
The working fluid is $^{4}He-II$
and so the quantum of circulation has the 
value $\kappa= 9.97 \cdot 10^{-4} cm^2/s$. The calculation
is done at $T=1.3 K$ for which the other parameters 
of the problem have the values: $\nu= 23.30 \cdot 10^{-4} cm^2/s$
(with the ratio $\frac{\kappa}{\nu}= 0.42$),
$\rho_n=6.5 \cdot 10^{-3} g/cm^{-3}$, 
$\rho_s=138.6 \cdot 10^{-3} g/cm^{-3}$
(with $\frac{\rho_s}{\rho_n}= 21.323$),
$h=0.978$, $h_{\times}=4.0937 \cdot 10^{-2}$,
$h_{\times \times}=2.175 \cdot 10^{-2}$,
$d_{\times}=-2.045 \cdot 10^{-2}$ and
$d_{\times \times}=4.270 \cdot 10^{-2}$.

We employ periodic boundary conditions.
Both fluids are advanced with the 
same timestep. Provision has been taken so that the latter 
resolves adequately the fastest Kelvin waves present in the tangle.
This requirement leads to rather constricted time steps,
$\Delta t=0.483 \cdot 10^{-3} s$, which are of the order
of the viscous time scale in the normal fluid. 
The grid size for the Navier-Stokes calculation is $64^3$.
In this way the width of a numerical cell
is $\Delta x= l_{b}/64= 1.56 \cdot 10^{-3} cm$ and this length is used
also for discretizing the vortex loops. Here, $l_{b}=0.1 cm$
is the size of the system. Initially there are 
$351$ randomly oriented vortex rings with radii between $0.34 l_b$
and $0.45 l_b$ (and therefore with curvatures $c$ between 
$20 cm^{-1}$ and $30 cm^{-1}$). 
The initial tangle length is $L=77.9 cm$ and 
$N=99527$ vortex points are used for its discretization.
At the same time, the vortex line
density is $\Lambda=L/l_b^3= 0.779 \cdot 10^5 cm^{-2}$ and the average
intervortex spacing is $\delta \approx \Lambda^{-1/2}=0.0036 cm$
which corresponds to wavenumber $k_{\delta} \approx 277 cm^{-1}$. Our
vortex line density is representative of experimental conditions.
For example, the value reported in fig.2 of \cite{stalp:1999}
for grid velocity $v_g = 5 cm/s$ and $T=1.5 K$ 
was $\Lambda \approx 2 \cdot 10^5 cm^{-2}$. Our vortex line density
is larger than the value of $\Lambda \approx 0.18 \cdot 10^5 cm^{-2}$
(again for $v_g = 5 cm/s$ and $T=1.5 K$), reported in the same experiment 
at saturation when the observed classical decay begins.\\
We calculate the average kinetic energy $E_n$, enstrophy $\Omega$ and
helicity $H$ of the normal fluid defined as:
$E_{n}=\frac{1}{2 V}\int \boldsymbol{u}\cdot\boldsymbol{u}\, 
d\boldsymbol{x}$, $\Omega= 
\frac{1}{2 V}\int \boldsymbol{\omega}\cdot\boldsymbol{\omega}\, 
d\boldsymbol{x}$ and $H= 
\frac{1}{2 V}\int \boldsymbol{u}\cdot\boldsymbol{\omega}\,d\boldsymbol{x}$.
In these relations $\boldsymbol{u}$ is the fluctuating
part of the normal fluid velocity and $V=l_b^3$ is the system volume. 
Using $E_n$ we can define a $Re$
number for the normal fluid velocity fluctuations: $Re= {\mathtt u} l_b / {\nu}$
where ${\mathtt u}=\sqrt{\frac{2}{3} E_{n}}$ is the intensity of the velocity 
fluctuations. We also calculate the normal fluid velocity spectrum $E_{n}(k)$ 
having the property: $\frac{1}{2 V}\int \boldsymbol{u}\cdot\boldsymbol{u}\,
d\boldsymbol{x} = \int_0^{\infty} E_{n}(k) dk$.

The results of Fig.\ref{diagsone} show that (due to excitation from 
mutual friction force) energy is being transfered from the 
superfluid to the normal fluid with simultaneous decrease in
vortex tangle length. The latter does not occur in a monotonous fashion.
In particular, around $t=0.03 s$ a local enstrophy maximum 
(consistently associated with a reduction in the slope of normal kinetic energy growth)
is accompanied by rapid vortex length growth.
This phenomenon could be explained by noticing
that the initial vortex configuration is not the natural
state of the system. There must be a transient length increase 
having to do with the induction
and evolution of (reconnection triggered) Kelvin waves along the vortices.
Indeed since the time step resolves the fastest Kelvin waves and until $t=0.03 s$
approximately $150$ time steps were taken,
adequate time was available for vortex wave growth.
In this milieu, the previously mentioned enstrophy peak
could be a manifestation of the established \cite{kivotidesr:2001}
intensification of dissipation rate in the normal fluid
due to reconnection associated  Kelvin waves.
The plausibility of this explanation was further supported by observing the actual
configuration of the tangle. Overall, the
recorded tangle length growth is a systemic transient towards a generic vortex
configuration. In this generic state the reconnections number 
sustains a linear growth of constant slope (Fig.\ref{diagsone}). Moreover,
although there are initially $351$ loops in the system,
at the end of the transient there are approximately $10$, a number that 
remains constant afterwards with $2-3$ loops having more than $90\%$
of the tangle length.\\
The computation also shows that (at $t\approx 0.097 s$) 
there is a critical tangle length $L_{c}=74.55 cm$ and 
an associated fractal dimension (measured with the
box counting algorithm \cite{kivotidesf:2001}) $D_{c}=1.87$ for which the 
normal fluid energy attains a peak. The fractal scaling was observed over
almost two decades from the system size ($l_{b}=0.1cm$) to the
discretization length along the vortices ($\Delta x= 1.5 \cdot 10^{-3}$).
This could mean that as the length decreases and the normal fluid 
volume not in the vicinity of a quantized vortex increases,
regions appear in the normal fluid where the viscous action is 
not counteracted by mutual friction excitation. In these
regions energy can only be dissipated into heat. 
Notice that at this time $86\%$ of the tangle's length belonged to a single
loop. The latter loop was also found to be a fractal with dimension $D_{gl}=1.84$.\\
Helicity keeps oscillating around zero. Since helicity is identically zero
for two dimensional vortical flow, this diagnostic might be an indication of a tendency of 
the normal fluid flow to occur (locally) on planes normal to the superfluid vortices
and the normal flow vorticity (see \cite{kivotidess:2000} for a demonstration of this).\\
The velocity spectrum at normal kinetic energy peak is found in Fig.\ref{diagstwo} to
scale like $E_{n}(k) \propto k^{-2.2}$. 
The end of the energy spectrum coincides with the average intervortex spacing
wavenumber (for the same time) $k_{\delta} \approx 273 cm^{-1}$. Comparing 
with $E_{n}(k) \propto k^{-5/3}$
of inertial turbulence, we comment that the present flow does not
comply with the familiar classical turbulence concept
of energy injection at large scales
and its subsequent transfer by nonlinear terms
towards smaller scales through a local in spectral space cascade.
Instead, normal fluid motion is being simultaneously excited and dissipated
at all resolved flow scales
by mutual friction and viscous forces respectively.
The curvature histogram $h(c)$
in Fig.\ref{diagstwo} (again at energy peak) which indicates that reconnections
exite in the initial tangle curvature scales much finer
than the resolved velocity scales supports this argument.
The acquired energy would tend to remain localized in the respective
wavenumber space regime where it first appeared. This is because
energy fluxes in wavenumber space are nonlinear effects
\cite{frisch:1995} and the present nonlinearity in the 
normal flow is weak (since if it was strong
we should have seen the formation of an inertial range with high $Re$ number
in the normal fluid).\\
It is important not to confuse the present turbulence with the dissipation
regime of classical turbulence. Although 
viscous effects are strong and the spectrum slope is steep, at the same time
the system is forced at all resolvable scales and we do not
have an exponential cut-off. In fact, the recorded $-2.2$ scaling exponent
is less steep than the $-3$ energy scaling exponent of the direct inertial
enstrophy cascade in two dimensional turbulence \cite{lesieur:2001}.
Incidentally, the energy flux in the direct enstrophy cascade of the 
latter case is also zero \cite{lesieur:2001} holding a resemblance
to the present case.\\
From a somewhat different perspective,
one notes that in the computation of \cite{kivotidess:2000} involving the 
propagation of a single superfluid ring,
the circulation strength $\kappa_n$ of the induced normal vortices
was found to be of order $\kappa$. For $^{4}He-II$ $\kappa$ is of the order of $\nu$.
For example, at $T=2.171$ when $\frac{\rho_s}{\rho_n}=0.0467$ it is $\frac{\kappa}{\nu}=5.47$
which is an indicative upper limit for this parameter. For comparison, in classical fluids
$\frac{\kappa_n}{\nu}$ could easily acquire values of the order $10^6$. Because of these,
the induced normal flow in \cite{kivotidess:2000} was a highly dissipative flow.
The present results suggest that the above physical picture is not affected by the much
higher vortex line density. It is obvious that vortex tangles much denser 
than those found in \cite{stalp:1999} are required in order to
achieve induced normal flow Reynolds numbers similar to those found in classical
fluid dynamics.\\

Overall, the present data lead to a number of conclusions about 
high Reynolds number quantum turbulence.\\
First, one can conclude that in fully developed $^{4}He-II$ 
turbulence the normal fluid inertia
is due to imposed pressure gradients and external
stirring (e.g. by towing a grid) rather than to excitation from superfluid vortices. 
Definitely, the present calculation does not exclude 
the possibility that at high
$Re$ number the normal flow might introduce a kind of organization
in the vortex tangle, for example 
bundles of aligned quantized vortices functioning
like classical ones at large enough scales.
Then, the superfluid vortex tangle could in turn 
stir vigorously the normal fluid (at these large scales).
However, even if something like this does happen,
it ought to bridge the gap between the normal fluid $Re$ number 
of order $1$ of the present calculation and the normal fluid $Re$ numbers
$10^3< Re <2 \cdot 10^5$
of the experiment of \cite{stalp:1999} (with analogous to ours
vortex line density and temperature) or $Re= 1.4 \cdot 10^5$
in the experiment of \cite{tabeling:1998}. By the same token, 
the assumption of \cite{stalp:1999}
that the two fluids have comparable vorticities seems unlikely.\\ 
Second, there are cases where the action
of the mutual friction force could cause an
initially laminar normal flow to become unstable and subsequently turbulent
\cite{melotte:1998}. This computation suggests that the ensuing
normal fluid turbulence would still be maintained
by interaction of the normal fluid Reynolds stresses
with normal fluid mean velocity gradients
induced by the instability \cite{schlichting:1999} and not
by (the meager) direct energy input from the superfluid vortices.\\
Third, (and in agreement with the discussion
of \cite{kivotides:2002}), in high $Re$ number 
quantum flows of previous experiments \cite{tabeling:1998,stalp:1999} the
observed energy spectra should have been to a very good approximation
the unforced Navier-Stokes spectra of the normal fluid. This conclusion could be reached
as follows: (a) in the \cite{stalp:1999} experiment,
the normal fluid spectrum is the dominant one. To prove this, we first notice that 
the superfluid kinetic energy in the latter experiment
should have been of the order of magnitude of the present one since the peak vortex line
densities are close. In the present calculation however, the superfluid energy 
is of the order of magnitude of the normal fluid energy
since the latter is due entirely to the presence of the quantized vortices.
By (safely) extrapolating from the current data, we conclude that even if
all initial tangle length was instantly transformed to normal fluid energy,
we could not match normal turbulent Reynolds numbers of order $10^4$
typical of experiments. Hence, statement (a) follows.
In order for the dominant normal fluid spectrum to be also of the unforced Navier-Stokes type,
it must be true that: (b) the magnitude of the mutual friction force
is much smaller than the magnitude of the normal fluid inertial terms.
This appears to be the case in \cite{stalp:1999} since if the mutual friction
force was comparable to the inertial terms one would have 
observed a vigorous energy transfer from the normal fluid to the superfluid.
In stating this, we take into account that according to proposition (a) 
the normal fluid energy exceeds significantly the superfluid energy.
This energy transfer would have resulted in an equally vigorous (by orders
of magnitude) growth of the 
vortex tangle length. Yet, the results of \cite{stalp:1999} show that exactly the opposite
happens (the length decreases). Therefore, in the experiment of \cite{stalp:1999}
mutual friction effects do not scale with inertial normal fluid effects
and the quantum fluid spectrum would be approximately 
that of the unforced Navier-Stokes dynamics of the normal flow.\\

The above conclusions could have been modified in case the turbulence 
in \cite{stalp:1999} was not decaying.
A key variable is vortex line density and the latter might not
attain in \cite{stalp:1999} high values because the normal fluid fluctuations 
(responsible for its growth at the initial rapid transient in the 
experiment) decay fast. 
In case a stationary normal fluid turbulence 
could be established via a sort of forcing, it would be possible for the
vortex tangle to reach a kind of equilibrium with the normal turbulence
characterized by a vortex line density corresponding to superfluid
energies comparable to those of the normal fluid. Notice that 
simply increasing the turbulence Reynolds number in \cite{stalp:1999}
might not have the latter effect since in this case the increase in superfluid energy
(denser tangle) would come together with high values of normal fluid turbulence
and the imbalance between the two could be preserved.
The associated complexity of such quantum flows
makes unlikely their calculation before the satisfactory 
resolution of a number of computational issues.
\begin{acknowledgments}
This research was supported by the Commission of the European Union
under Contract \# HPRI-CT-1999-00050. I thank Caltech for computing
time.
\end{acknowledgments}
\bibliography{dense}
\end{document}